# MECHANISM OF THE PHASE CHANGE IN PbK$_2$LiNb$_5$O$_{15}$: Dielectric, structural, and Raman scattering studies

Y. Gagou[1], M. El Marssi[1], M.-A. Frémy[3], N. Aliouane[2], D. Mezzane[4] and P. Saint-Grégoire[5*]

[1] LPMC, Université de Picardie Jules Verne, 33 rue Saint-Leu, 80039 Amiens Cedex, France
[2] HMI, Glienicker Strasse 100, D14109, Berlin, Germany
[3] L2MP, UMR-CNRS 6137, Université de Toulon-Var, BP 132, F-83957 La Garde Cedex, France
[4] F.S.T.G.-F.S.S., University Cadi Ayyad Marrakech, Morocco
[5] University of Bangui, Bangui, Central African Republic

**Abstract**:

Experiments reveal that PbK$_2$LiNb$_5$O$_{15}$ which belongs to the tetragonal tungsten bronze family presents paraelectric and ferroelectric phases and a complex structural change between them. High and low temperature phases are of symmetry $P4/mbm$ and $Pba2$ respectively, so that this change is also of ferroelastic type. As presented here, crystallographic results hint at a displacive character of the ferroelectric ordering but show a more complex behaviour, with a clear order-disorder mechanism which accompanies the appearance of ferroelasticity. To complete our knowledge of this material, we have performed Raman experiments which exhibit a low frequency mode, but no clear soft mode is observed.

## I - INTRODUCTION

PbK$_2$LiNb$_5$O$_{15}$ belongs to the Tetragonal Tungsten Bronze (TTB) structure as described in our previous works [1,2,3,4]. We recently determined unambiguously that the phase change in this material occurs between $P4/mbm$ and $Pba2$ space groups, in good agreement with the behaviour of dielectric constant showing a lambda-shaped anomaly in the c-axis direction. The phase change reveals intermediate states or phases in the region of the transition, the study of which is in progress.

Because of this complex nature, we have performed a more detailed dielectric study, including the temperature evolution of dielectric relaxation in a wide temperature range. The dielectric results are presented in the third part. In the fourth part we present in detail results on the crystal structure at high and low temperatures, in paraelectric-paraelastic phase and in ferroelectric-ferroelastic phase. To check the displacive character which may be deduced from dielectric and structural data, we performed Raman scattering experiments, the results of which are shown in part five. All these results cleared up the physical properties of the material.

---

[*] Permanent address: 14, Av. F. Mistral, 34110 Frontignan, France
[*] E-mail: stgreg@ferroix.net

## II. – EXPERIMENTAL CONDITIONS

### II.1. Synthesis of samples

$PbK_2LiNb_5O_{15}$ powder is synthetized from oxides ($PbO$, $Nb_2O_5$) and carbonates ($K_2CO_3$, $Li_2CO_3$) of 99.99 % purity by solid state reaction according to the equation [5]:

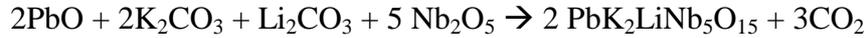

$$2PbO + 2K_2CO_3 + Li_2CO_3 + 5\,Nb_2O_5 \rightarrow 2\,PbK_2LiNb_5O_{15} + 3CO_2$$

We started from a mixture in stoichiometric proportions and ground it by ball milling in an aqueous solution during 24 hours. After the evaporation of water, the resulting powder was pressed under 4 bars to form cylinder-shaped pellets. These pellets were then annealed successively at 1200, 1270 and 1320 K with grinding and X-ray diffraction control between each annealing. After numerous syntheses and trials, this way has empirically appeared to give the best results, both for the chemical composition of the final product and its crystalline quality. The single crystals used for dielectric measurements were prepared by the flux flow method at 1423 K, with an excess of $PbO$ and $B_2O_3$. Powders obtained by crushing them were shown to have the same features as powders synthetized according to the above method.

### II. 2. Dielectric measurements

The investigated sample is a $PbK_2LiNbO_{15}$ platelet-shaped single crystal with large surfaces of 2,465 $mm^2$ area and a thickness of 0,985 mm ; the (silver paste) electrodes covered the large areas of the sample, oriented perpendicular to c-axis. The experiments were performed in the same furnace that was used during dielectric measurements (see ref.[2]). The investigated temperature range is 300K- 800 K.

The dielectric measurements were performed using a HP 4284A impedancemeter as in our preceding study [2]. This apparatus has a sensitivity of 0.05 %, and allows to determine accurately the variations of the sample electrical impedance. Impedance spectroscopy studies were carried out by means of a SOLARTRON SI-1260 spectrometer in the frequency range from 5 to 32 MHz. Measurements were performed using a source of 1Vrms in four points configuration. We displayed the results in Nyquist diagrams and deduced the relaxation time parameters.

### II.3. Structural studies

The experimental conditions and techniques are the sames as those described in another paper dealing specifically with the crystallographic study itself [6]. We use here the final results obtained in this preceding work where the atomic positions were determined for both phases -at high and low temperature-, but were not deeply analyzed.

X-ray diffraction data on crystalline powders were collected at 290 K and 670 K, using the high resolution mode of the *DW*22 instrument at the synchrotron laboratory LURE close to Paris. The used wavelength $\lambda = 0.69181$ Å, selected with a double Si (111) monochromator, was chosen for optimizing conditions, taking into account absorption, flux and resolution.

Data were recorded using the Debye-Scherrer geometry in an angular domain between 5 and 60° in $2\theta$ with steps of 0.004°. The powder sample was in a silica capillary tube of diameter 0.5 mm, heated with a hot gas flow, the stability at high temperature being around 1 K.

### II.4. Raman spectroscopy measurements

Raman spectra were measured in a backscattering microconfiguration, using the polarized light of an argon ion laser (514.5 nm) focused on the sample surface as a spot of 2 µm diameter. The scattered light was analysed using a Jobin-Yvon T64000 spectrometer equipped with a CCD. The spectra were recorded on the single crystal in $Y(ZX)\bar{Y}$ polarization where Z is along the c-axis and X, Y are along the (a, b) axes of the crystal, respectively. All the reported spectra have been corrected by the Bose-Einstein factor. The temperature variation during Raman experiments was performed using a Linkam TS 1500 hot stage allowing a temperature stability of ± 0.1 K.

## III. DIELECTRIC STUDY

The temperature dependence of the inverse of dielectric constant is shown in Fig.1: as measured in the c-crystallographic direction, the dielectric constant shows a λ–shaped curve, and we can notice the existence of a thermal hysteresis and a stepwise behaviour in the corresponding region, which shows that there exists a first order phase transition in the system. Note that the observed thermal hysteresis has an anormal character since it is prolounged in the low temperature phase far away from the transition region : curves obtained on cooling and on heating coincide approximately 150 K below the temperature at which the jump occurs. Even if such a situation is not frequent, it is not however a single case : in the analogous compound $Ba_2NaNb_5O_{15}$, a large global hysteresis is also present. Detailed studies revealed the origin of this phenomenon as due to the presence of an incommensurate phase and to out-of-equilibrium states related to it [7]. In the case of the present compound, it is too early to give an explanation of this phenomenon, which requires further studies ; it seems nevertheless possible that the extension of the hysteresis may be related with the presence of intermediate phases or states in the region of the transition, conferring a complex nature of the structural change.

The inverse of dielectric constant versus temperature shows that a Curie behaviour is followed in a wide temperature range but in the close vicinity of the transition, there appears a deviation (see the inset of Fig. 1). The value of Curie constant that can be deduced is C = 1,5×10$^5$ K. This value is close to values characterizing such displacive ferroelectrics as $BaTiO_3$ and $LiTaO_3$, as described by Jona and Shirane (1962) [8, 9, 10].

Let us now present results obtained in a wider frequency range (5-10$^6$Hz) by impedance spectroscopy measurements. These experiments allowed us to display the results at different temperatures in Nyquist diagrams. From them, we analyzed the relaxation, and a relaxation time can be deduced at each temperature. Only one relaxation time is evident in this sample : all the obtained Nyquist diagrams are indeed half-circles centred on Z' (real Z) axis.

We present in Fig.2 the temperature dependence of the system relaxation frequency, where each point is extracted from Nyquist diagram adjustment. The relaxation frequency $f_p$ (inverse of the relaxation time) shows a strong temperature dependence, with an abrupt decrease on approaching the temperature $T_C$. Mention also that the temperature dependence observed in Fig.2 presents a complex behaviour, consistent with the known complex character of the structural change.

# IV. STRUCTURAL STUDY : COMPARISON BETWEEN HIGH AND LOW TEMPERATURE PHASES

The crystallographic structure of the (tetragonal) paraelectric phase was refined using the Rietveld method available in the program *Fullprof* [12]. The cell parameters in this phase were determined as $a_t = b_t = 12.6495(4)$ Å and $c_t = 3.9719(4)$ Å. The refinement of the lattice constants, atomic coordinates, isotropic atomic displacement parameters and chemical occupancies led to a rapid convergence.

The best fit was achieved by refining the anisotropic atomic displacement parameters for the Pb atoms, and final results were presented by introducing one lithium atom per formula unit (so that half of triangular sites is occupied). All the site occupation rates had been determined [6]. Thus we focus our study on the atomic coordinates changes through the phase transition.

## IV.1. Change of unit cell and Atomic positions in both phases.

Results of the structural study previously obtained [13] are given for both phases *P*4*/mbm* and *Pba*2 in Tables 1 and 2 respectively. Structure refinement in the tetragonal space group *P*4*/mbm* is satisfactory with a remarkable agreement between the result of the calculated diagram and the experimental profile. Fig.3 presents the Fourier map that has been drawn using observed structure factors and calculated phases [13]. We present in the same figure the unit cells of this structure in the two orthorhombic (*Cm*2*m* and *Pba*2) space groups previously considered for describing the low temperature phase of this compound, to better allow to understand the differences between both structures.

Since according to our previous studies, the ferroelectric and ferroelastic phase belongs unambiguously to the *Pba*2 space group [6, 14], we use in the remaining part of this paper the atomic positions obtained with *Pba*2. To understand the mechanism of the ferroelectricity, it is necessary to compare the atomic positions in the groups *P*4/*mbm* and *Pba*2, and more particularly along the c-axis which is the polar axis.

This has been achieved by means of the Ortep software [15]. In Fig.4, we represent the crystal structure along the c-axis of $PbK_2LiNb_5O_{15}$ using the representation of atomic thermal displacements spheres in the ferroelectric phase (left) and in the paraelectric phase (right).

The three types of sites [octaedral (Fig.4-a), square (Fig.4-b) and pentagonal (Fig.4-c)] are well resolved. One can see a niobium atom (small gray sphere at interconnection of oxygen atoms) in octahedral sites. Square and pentagonal tunnels are occupied by lead (big lighted gray spheres) and potassium (big black spheres) atoms, in the proportions 54.4/45.6 % and 23.6/76.4 % [13] respectively. Thus the triangular sites are assumed to be occupied by lithium ions (not shown).

Both structures differ weakly at a first glance, though change of positions associated with ferroelectricity and with ferroelasticity are clearly seen. However, it is obvious that the features of disorder are different in both phases.

## IV.2. Site configurations

It is useful to analyze the mechanism of the structural change by focusing on the change of position of the "heavy ions" $Pb^{2+}$, $Nb^{5+}$, $K^+$, in the different sites at high and low temperatures. The lithium will not be considered since its contribution to the structure factor is very weak and it is not evidenced by x-ray diffraction. However, during the structural refinements, we imposed the lithium position in the triangular sites of the structure, which seems reasonable. In Table 3, we list the atom coordinates and displacements, in the three space directions for the octahedral, square and pentagonal sites. Static displacements of atoms are calculated by taking into account the coordinates in the paraelectric and ferroelectric phases.

a) Octahedral sites

We present in Fig.4 (a) the position of niobium atoms in the octahedral site and the connections with the neighbouring oxygen atoms in the ferroelectric (left) and paraelectric (right) phases. In ferroelectric phase, one notices a slight distortion, with the atom Nb2 (see Table 1 and Table 2) slightly out of the equatorial plane of the $NbO_6$ octahedra, resulting in an angle (O-O-Nb) ~ 4° and a niobium displacement of 0.0237Å along c-axis. In addition, Table 3 shows that, from the octahedron mediator plane perpendicular to c-axis, the niobium atoms Nb2 undergo displacements in the three crystallographic directions. These displacements are of antiferroelectric nature in the (a, b) plane. Notice that their positions are connected with the breaking of tetragonal symmetry, namely ferroelasticity. Along the c-axis, the displacements have a polar symmetry and are thus related to the ferroelectricity. The result may be consistent with a displacive mechanism : the thermal displacement spheres are similar in both phases, and nothing may suggest an order-disorder mechanism related to this type of atoms : the atomic displacement related to the structural change occurs within the thermal displacement sphere.

b) Square sites

In the square sites (2 per unit cell), the phase change seems to have also a character rather displacive, with the displacement of $K^+$ ions in the c-axis direction [Fig.4 (b) (left)]. The change of atom positions is related to ferroelectricity : one sees, according to the Table 3, that $Pb^{2+}$ ions as well as $K^+$ ions are moved in the c-direction, and thus contribute to the polarisation. On the other hand, the thermal displacement spheres are similar in both phases, and in particular the thermal displacement sphere at high temperature does not contain that at low temperature. Surprisingly (since that result, ferroelectricity was considered as induced by $Pb^{2+}$ ions in this family of compounds), the $K^+$ ions bring a larger contribution to polarization. One sees again, according to Table 3 and Fig.4 (b), that the ferroelectric change may be induced by a progressive shift of atomic position without big changes of atomic displacement parameters: there is no evidence of any order-disorder mechanism.

c) Pentagonal sites

In the pentagonal sites (in number of four per unit cell), one observes (Fig.4-c) that the change is primarily related to a rearrangement of the atoms in the (a, b) plane. The nature of these displacements is antiferroelectric : the $K^+$ ions in central position in the high temperature phase are located, at low temperature, in the vicinity of the equatorial plane. This leads to a dipolar moment perpendicular to c-axis which is compensated by the configuration in the other sites (resulting from the symmetry operations of *Pba*2 space group). Thus, in the unit cell, the component of the polarisation ***P*** in the (a,b) plane is averaged to zero. Again, there is no evidence of order-disorder mechanism for $K^+$ ions. However concerning $Pb^{2+}$ ions, there are in principle two possibilities at high temperature. The first one is that the two positions found are occupied statistically with the same probability, and in a dynamic way (for a given site). The second possibility corresponds to a static disorder with ions in both types of sites.

At this stage of our structure study we cannot distinguish both possibilities (static or dynamic) when considering only one phase. Mention nevertheless that the comparison of structural data at high and low temperatures is consistent only with the hypothesis of dynamical occupation at high temperature, with a change of probability of site occupation as a function of temperature, which is the classical scheme for the order-disorder transition. The alternative explanation would involve diffusions of atoms on distances larger than the unit cell, reversible with temperature, which is an irealistic process.

### IV.3. Order parameters and changes of atomic positions

In the following, we work using the assumption that the occupation of the two sites is of dynamic nature. In Fig.4-c, one notices that the $K^+$ and $Pb^{2+}$ ions are at the same height along c-axis (see Table 3) but the ellipsoid of atomic displacements is larger for $Pb^{2+}$ than for $K^+$. Table 3 shows an overall change of localisation of the ions $Pb^{2+}$ and $K^+$ along x, y and z axis. Concerning the $Pb^{2+}$ ions, there are 2 sites occupied at high temperature [see Fig.4-c (right)], and only one site at low temperature [(see Fig.4-c (left)]. The latter coincides practically with one of the two sites characterizing the high temperature configuration : we are clearly in the order-disorder situation. The $Pb^{2+}$ ions thus take part in an order-disorder mechanism in the (a, b) plane. Nothing similar occurs in the c direction.

Displacements in the (a,b) plane create a distortion of the unit cell associated with the breaking of tetragonal symmetry, namely with the appearance of ferroelasticity.

Mention that the changes of atomic positions inducing ferroelasticity are by no means coupled to those inducing ferroelectricity : the atomic displacements along z do not induce the change of atomic positions related with ferroelasticity. In other words, neither the polarisation nor the lattice distortion is a secondary order parameter of the transition. This case distinguishes itself also clearly from ferroelectric and ferroelastic phase transitions, a typical example in the tetragonal system being with a polarization lying perpendicular to the high symmetry axis : in such a case the polarization itself breaks the tetragonal symmetry and then induces also ferroelasticity.

Here at the opposite, the nature of changes of position leading to both different orders (ferroelectric and ferroelastic) indicate that two order parameters (in the sense of Landau and Lifshitz) condensate in the low temperature phase. This is a consequence of the following : symmetry properties of "ferroelectric" and "ferroelastic" position changes are different and the appearance of one ordering is not the consequence (in terms of symmetry breaking) of the other.

This result is consistent with the data that may be found in tables on possible symmetry changes at structural phase transitions [18, 19] where no irreducible representation of *P4/mbm* may induce a transition to *Pba*2 with the cell correspondance we have in our case.

Crystallographic results show that one type of change of atomic position (leading to ferroelectric order) could be of displacive nature, whereas the other one (leading to ferroelasticity) is clearly of order-disorder type.

# V. RAMAN SCATTERING STUDY

In the orthorhombic ferroelectric phase ($C_{2v}$ point group) all the optical modes $A_1(z)$, $A_2$, $B_1(x)$, and $B_2(y)$ are Raman active. The coordinates in brackets denote the direction of the dipole moments and x, y, z refer to the principal axes of the orthorhombic phase. In the case of the $A_1$, $B_1$, and $B_2$ polar modes, longitudinal (LO) and transverse (TO) characters should be considered.

For $D_{4h}$ point group associated of the tetragonal paraelectric phase, all the modes $A_{1g}$, $B_{1g}$, $B_{2g}$, and $E_g$ are Raman active. $E_g$ mode is doubly degenerate.

Temperature dependence of Raman spectra recorded in different geometries do not show any soft mode in the $C_{2v}$ phase when the temperature approaches the critical one, from below. However in the TTB compounds it was shown that the ferroelastic-ferroelectric phase transition is characterized by a splitting of the E modes of the tetragonal ferroelastic phase into two components $B_1$ and $B_2$ of the orthorhombic ferroelectric phase [16, 17]. Therefore we present in this work, a limited area of the Raman spectra showing only the splitting of B modes.

Fig.5-a displays the temperature dependence of the low frequency part of the $Y(ZX)\bar{Y}$ spectra. Note that the X and Y directions are included in the (a, b) plane of the crystal. Therefore the $Y(ZX)\bar{Y}$ spectra exhibit a superposition of $B_1$ and $B_2$ modes in $C_{2v}$ phase. As the TTB compounds, the very low frequency (25 - 82 cm$^{-1}$) can be attributed, from low to high frequency, to the $B_2$ and $B_1$-$B_2$ pair degenerated from $E_g$ mode of paraelectric tetragonal phase.

In this frequency range, the Raman spectra can indeed be fitted using three lines (47.16, 59.20 and 66.44 cm$^{-1}$) as can be shown in Fig. 5-b at room temperature. At the difference of TTB compounds (BNN for example), the frequency of the $B_2$ low frequency mode does not change when the temperature increases. However, close to the transition temperature, Raman spectra show a dramatic change as shown in Fig. 5-a. The lowest part of frequencies changes to one large band which can be attributed to $E_g$ mode of $D_{4h}$ phase. This change can be explained by the disappearance of the $B_2$ mode and the change of $B_1$-$B_2$ pair to $E_g$ mode. We attributed the large width of the $E_g$ mode in the paraelectric phase to the disorder of the lead atoms in the pentagonal sites, as explained above. Furthermore, the $B_1$-$B_2$ modes located at 140 cm$^{-1}$ in the ferroelectric phase disappear in the paraelectric phase.

Observation of doublet splitting of Raman spectra may be an indication for superlattice formation, but we did not consider this possibility because it is not supported by structural results, neither in present nor in preceding structural studies (including transmission electron microscopy and electron diffraction) [2, 3, 6, 13, 14]. In particular, electron diffraction is a powerful tool for evidencing superlattice reflections, but no one of such reflections is observed in electron diffractograms. We therefore consider the low temperature structure as shown in the present figures as valid.

## VI- CONCLUSION

This work, which permitted us to identify $PbK_2LiNb_5O_{15}$ as a specific ferroelectric distinguishing itself from the others, led us to look further into its properties.

After determining the crystallographic structure in the two phases (high and low temperatures), we focused on the details of the structural arrangement and on the difference between phases. It appears that besides a possible displacive character associated with the ferroelectricity, an order-disorder mechanism is involved in the onset of ferroelasticity.
Raman experiments exhibit clearly differences between the ferroelectric-ferroelastic and the paraelectric-paraelastic phases. However no soft mode was observed, perhaps because of an overdamping of the ferroelectric soft mode. These results confirm that the involved mechanisms are not simple, and further studies are necessary to complete our understanding of this compound.

*This work was supported by NATO collaborative linkage grant PST.CLG.980055*

# Pictures and Tables

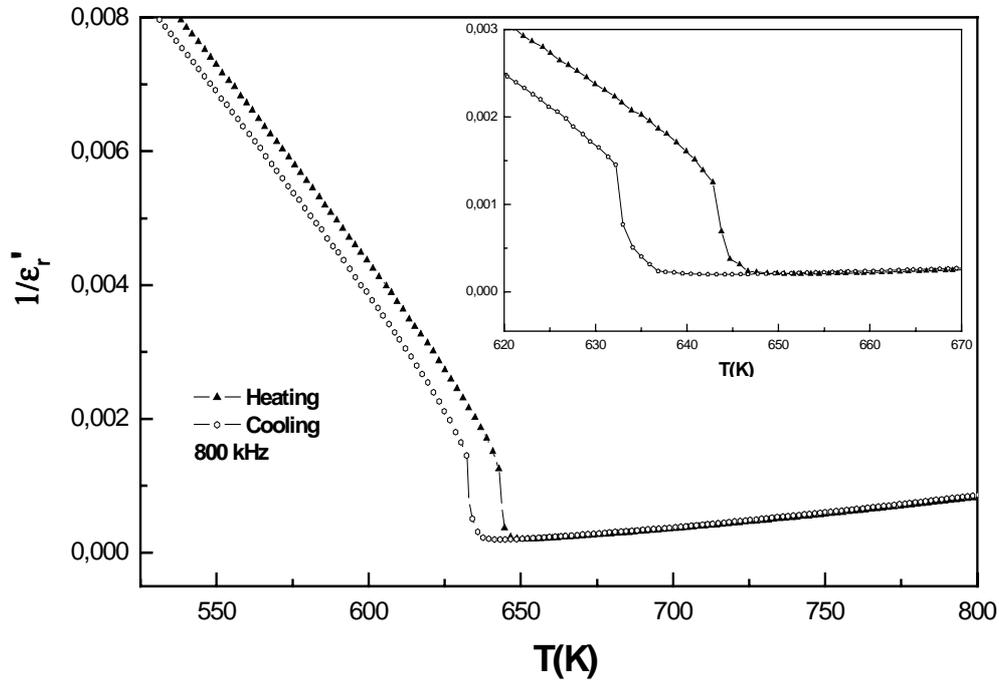

**Fig.1**

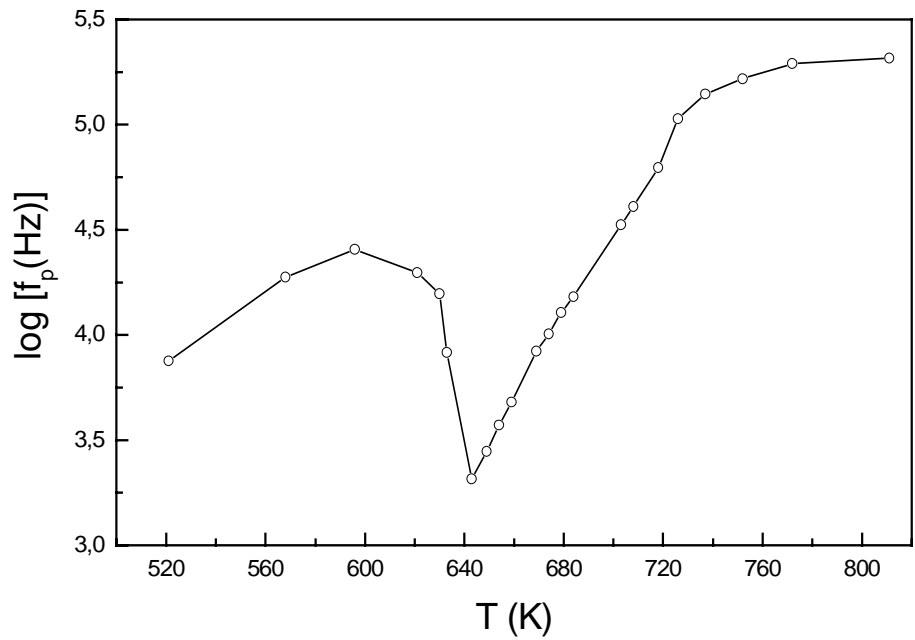

**Fig.2**

# Table 1

| Atoms | x | y | z | $B_{iso}$ (Å²) | Occup. | |
|---|---|---|---|---|---|---|
| Pb1 | 0,00000 | 0,00000 | 0,50000 | 2,53(4) | 0,069 | Square sites ($s_1$) |
| K1 | 0,00000 | 0,00000 | 0,50000 | 1,68(5) | 0,056 | |
| Pb2 | 0,1517(8) | 0,7045(8) | 0,50000 | 6,54(2) | 0,059 | Pentagonal sites ($p_2$) |
| K2 | 0,1624(3) | 0,6624(3) | 0,50000 | 3,24(9) | 0,191 | |
| Nb1 | 0,00000 | 0,50000 | 0,00000 | 1,69(3) | 0,125 | |
| Nb2 | 0,0749(2) | 0,2104(4) | 0,00000 | 1,40(9) | 0,500 | |
| Li | 0,6198(7) | 0,1198(7) | 0,50000 | 2,000 | 0,125 | Half of triangular sites are occupied |
| O1 | 0,00000 | 0,50000 | 0,50000 | 1,76(8) | 0,125 | |
| O2 | -0,0026(5) | 0,3430(5) | 0,00000 | 2,80 | 0,500 | |
| O3 | 0,0726(8) | 0,2107(5) | 0,50000 | 2,43(8) | 0,500 | |
| O4 | 0,2871(6) | 0,7871(6) | 0,00000 | 2,04(3) | 0,250 | |
| O5 | 0,1403(9) | 0,0674(1) | 0,00000 | 2,02(4) | 0,500 | |

# Table 2

| Atoms | x | y | z | $B_{iso}$ (Å²) | Occup. | |
|---|---|---|---|---|---|---|
| K1 | 0,00000 | 0,00000 | 0,4293(2) | 1,623 | 0,231 | Square sites ($s_1$) |
| Pb1 | 0,00000 | 0,00000 | 0,5056(8) | 1,623 | 0,269 | |
| K2 | 0,2056(9) | 0,6435(9) | 0,4988(8) | 5,078 | 0,743 | Pentagonal sites ($p_2$) |
| Pb2 | 0,1576(3) | 0,6792(1) | 0,5035(8) | 5,078 | 0,257 | |
| Nb1 | 0,00000 | 0,50000 | 0,00000 | 1,367 | 0,500 | |
| Nb21 | 0,07530 | 0,2108(5) | -0,0056(9) | 1,494 | 1,000 | |
| Nb22 | -0,2108(7) | 0,0746(8) | -0,0059(5) | 1,307 | 0,500 | |
| Li | 0,8773(9) | 0,6184(6) | 0,50000 | 1,0 | 0,500 | Triangular site 1 atom per formula unit |
| O11 | 0,3490(4) | 0,0005(1) | 0,0365(3) | 2,106 | 1,000 | |
| O12 | -0,00400 | 0,3401(5) | 0,0429(5) | 2,106 | 1,000 | |
| O21 | 0,1441(1) | 0,0685(9) | 0,0205(8) | 2,106 | 1,000 | |
| O22 | -0,0664(8) | 0,13820 | 0,0365(1) | 2,106 | 1,000 | |
| O3 | 0,2137(5) | 0,2865(2) | 0,0332(5) | 2,106 | 1,000 | |
| O4 | 0,00000 | 0,50000 | 0,5210(5) | 2,106 | 0,500 | |
| O51 | 0,2780(4) | 0,4327(2) | 0,5157(5) | 2,106 | 1,000 | |
| O52 | -0,4184(7) | 0,3030(5) | 0,5078(8) | 2,106 | 1,000 | |

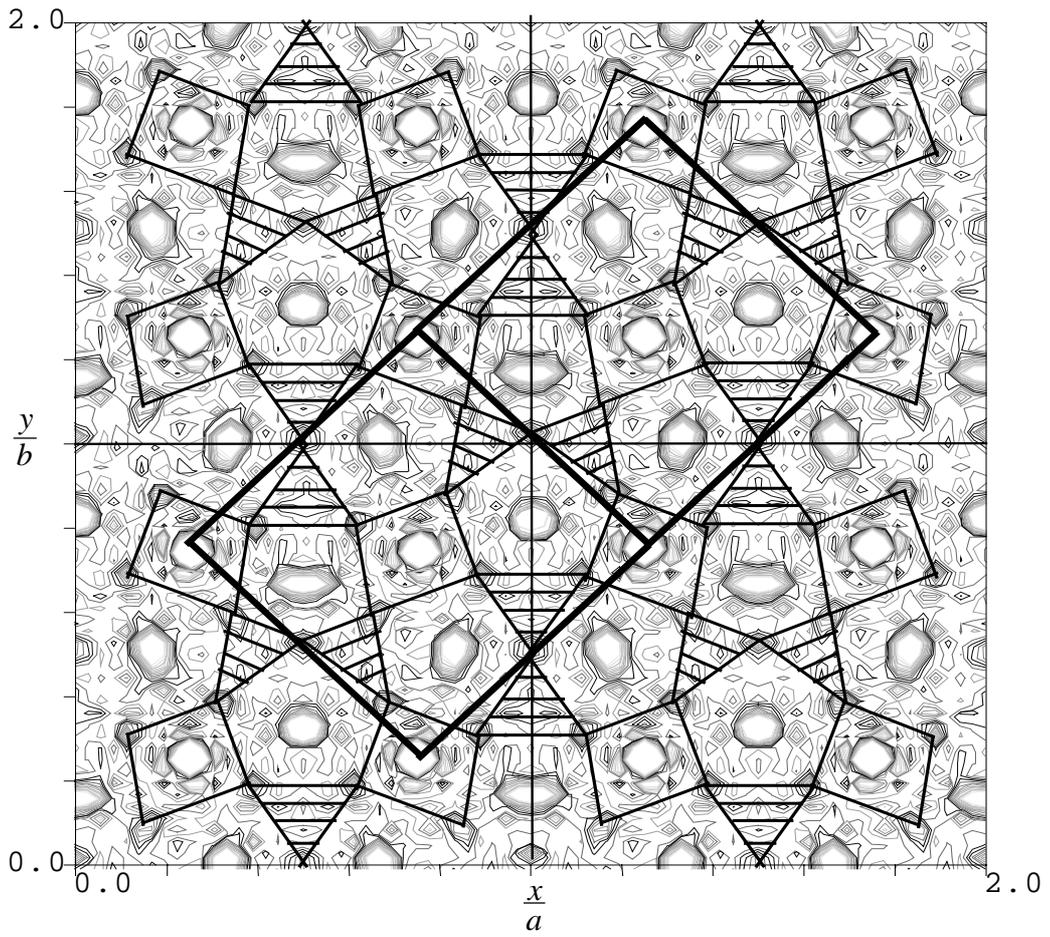

**Fig.3**

# Table 3

| Atoms | Axes | Pba2 | | P4/mbm | | Displ. |
|---|---|---|---|---|---|---|
| | | coord. (Å) | $u_{iso}$(Å) | coord.(Å) | $u_{iso}$(Å) | $\vec{d}$ (Å) |
| *Octahedral Sites* | | | | | | |
| **Nb** | <x> | 0.9459 | 0.0759 | 0.9474 | 0.0842 | -0.0015 |
| | <y> | 2.6613 | 0.0759 | 2.6614 | 0.0842 | -0.0001 |
| | <z> | -0.0237 | 0.0759 | 0 | 0.0842 | -0.0237 |
| *Square Sites* | | | | | | |
| **Pb(1)** | <x> | 0 | 0.0827 | 0 | 0.1038 | 0 |
| | <y> | 0 | 0.0827 | 0 | 0.1038 | 0 |
| | <z> | 2.0007 | 0.0827 | 1.9859 | 0.1038 | 0.0148 |
| **K(1)** | <x> | 0 | 0.0827 | 0 | 0.0817 | 0 |
| | <y> | 0 | 0.0827 | 0 | 0.0817 | 0 |
| | <z> | 1.6963 | 0.0827 | 1.9859 | 0.0817 | 0.2896 |
| *Pentagonal Sites* | | | | | | |
| **Pb(2)** | <x> | 1.9927 | 0.1464 | 1.9189 | 0.1668 | 0.0738 |
| | <y> | 8.5640 | 0.1464 | 8.9115 | 0.1668 | -0.3475 |
| | <z> | 2.2301 | 0.1464 | 1.9859 | 0.1668 | 0.2442 |
| **K(2)** | <x> | 2.5981 | 0.1464 | 2.0542 | 0.1226 | 0.5439 |
| | <y> | 8.1226 | 0.1464 | 8.3790 | 0.1226 | -0.2564 |
| | <z> | 1.9731 | 0.1464 | 1.9859 | 0.1226 | -0.0128 |

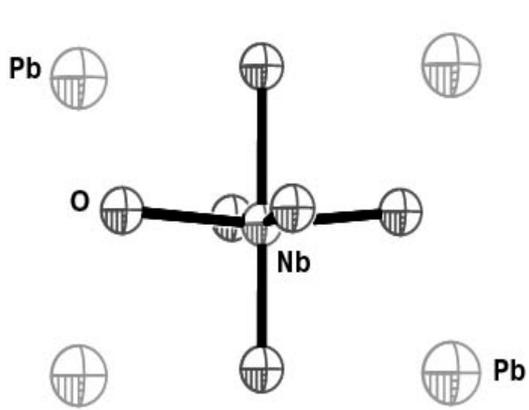
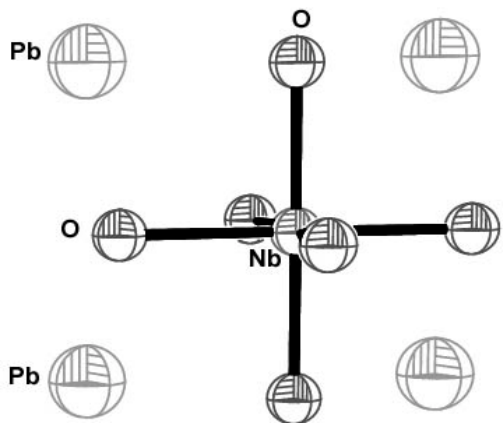

**(a) Octahedral sites**

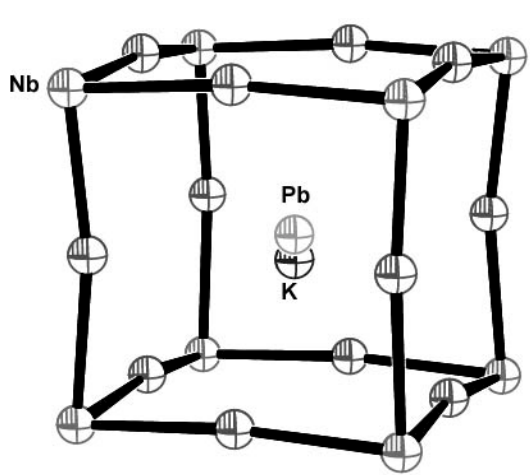
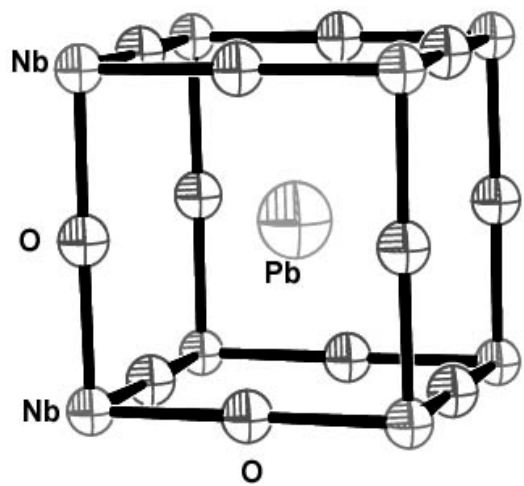

**(b) Square sites**

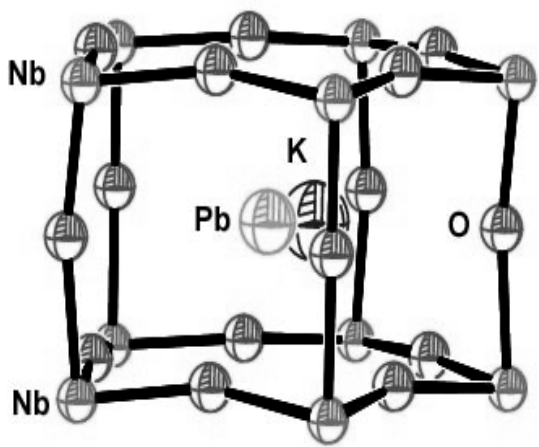
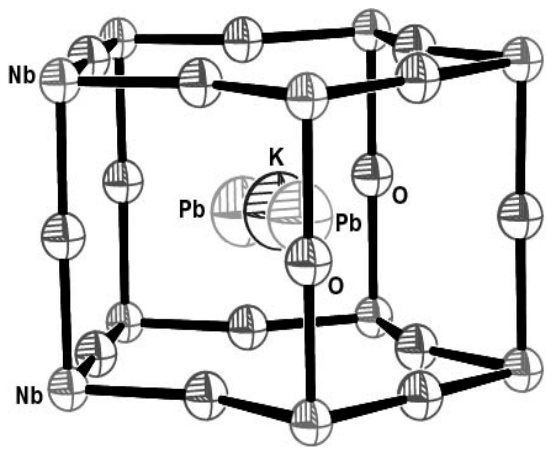

**(c) Pentagonal sites**

**Fig.4**

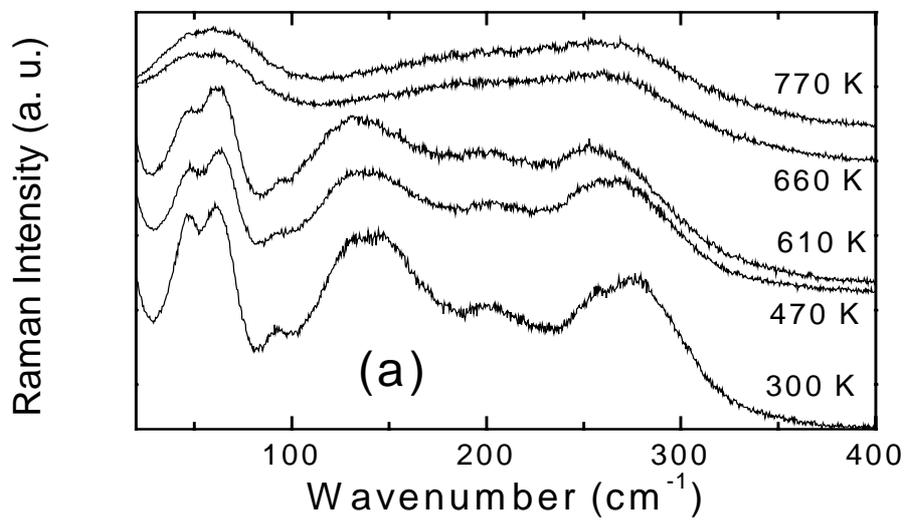
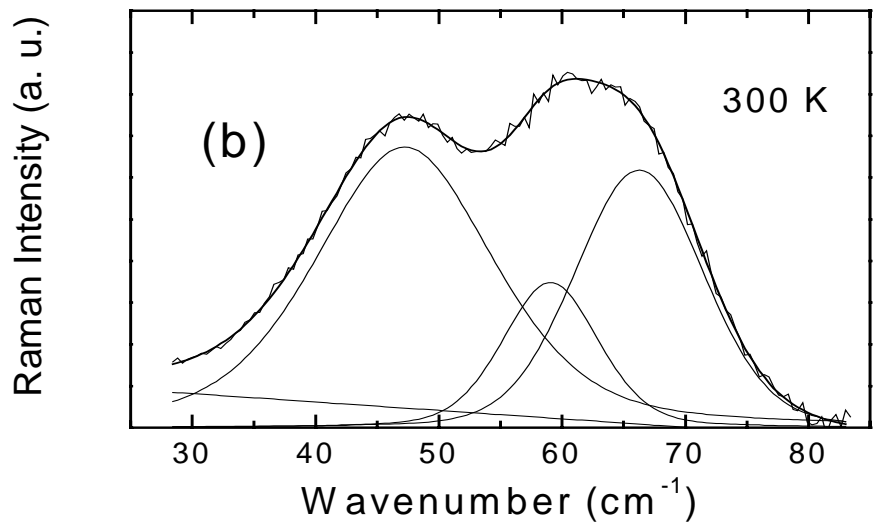

**Fig. 5**

# Figure captions

Fig.1: Inverse of the real part of dielectric permittivity versus temperature.
A Curie-Weiss law is followed except in the vicinity of the structure change. The inset shows the existence of a second anomaly above $T_C$.

Fig.2: Temperature dependence of relaxation frequency, showing the lowering around $T_C$

Fig.3: Fourier map at the height $z = ½$ exhibiting possible unit cells of the low temperature orthorhombic phase (in *Cm*2*m* and in *Pba*2), and two unit cells (bold lines) of *Pba*2 symmetry structure.

Fig.4 : Octahedral (a), Square (b) and Pentagonal (c) sites configuration in the ferroelectric and paraelectric phases : for each pair of figures the ferroelectric phase is on the left side and the paraelectric phase is on the right side.

Fig 5: (a) Temperature evolution of Raman spectra recorded in $Y(ZX)\bar{Y}$ geometry.
(b) Three-peak decomposition of Raman spectrum measured at room temperature showing the $B_2$ and $B_1$-$B_2$ modes.

# Table Captions

Table 1 : Atomic coordinates, thermal displacement parameters, and occupancy obtained from diffraction data in paraelectric phase (670 K), tetragonal with space group *P*4/*mbm*.

Table 2 : Atomic coordinates, thermal displacement parameters, and occupancy obtained from diffraction data in ferroelectric - ferroelastic phase (at 290 K), orthorhombic with space group *Pba*2.

Table 3: Atomic positions, isotropic thermal displacement factor and atomic displacement from mediator plane containing the barycenter G (chosen as reference) in paraelectric phase for octahedral, square and pentagonal sites.